\documentclass[aps,pra,groupedaddress,twocolumn,floatfix]{revtex4}

\newcommand{\ket}[1]{|#1 \rangle}

\newcommand{\pf}{\rm{^{40}K}} 
\newcommand{\re}{\rm{^{87}Rb}}

\usepackage{graphicx} 
\usepackage{amssymb}  

\begin{document}

\title{Rapid sympathetic cooling to Fermi degeneracy on a chip}

\author{S.\ Aubin, S.\ Myrskog, M.\ H.\ T.~Extavour, L.\ J.\ LeBlanc, D.\ McKay, A.\ Stummer, J.\ H.\ Thywissen}

\affiliation{Department of Physics, University of Toronto, Toronto, Ontario M5S 1A7, CANADA}

\date{\today}

\maketitle

%
\textbf{Neutral fermions present new opportunities for testing
 many-body condensed matter systems, realizing precision atom interferometry,
producing ultra-cold  molecules, and investigating fundamental forces.  However, since their first observation  \cite{Jin_Science1999}, quantum degenerate Fermi gases (DFGs) have continued to be challenging to produce, and have been
realized in only a handful of laboratories
\cite{truscott,SchreckDualD,thomas_fermi,Ketterle_Sympathetic_PRL2002,roati,%
InnsbruckDFG,ZurichDFG,ZimmermanRbLi,SengstockBECDFG}.
In this Letter, we report the production of a DFG using a
simple apparatus based on a microfabricated magnetic trap.  
Similar approaches applied to Bose-Einstein Condensation (BEC) of $\mathbf{^{87}Rb}$
\cite{Hansch_ChipBEC_Nature2001, Zimmerman_chipBEC2001}  
have accelerated evaporative cooling and eliminated the need for multiple
vacuum chambers.  We demonstrate sympathetic cooling for the first time in a
microtrap, and cool $\mathbf{^{40}K}$ to Fermi degeneracy in just six seconds ---
faster than has been possible in conventional magnetic traps.  To understand our sympathetic cooling trajectory, we measure the temperature dependence of the $\mathbf{^{40}K}$-$\mathbf{^{87}Rb}$ cross-section and observe its Ramsauer-Townsend reduction.}

Microfabricating the electromagnets used to trap ultra-cold atoms  
leads to a series of experimental benefits.
Decreasing the radius $R$ of a surface-mounted wire increases
the maximum magnetic field gradient
as $R^{-1/2}$ \cite{groth}. Since the oscillation frequency $\omega$ of the trapped atoms
increases linearly with transverse field gradient, decreasing $R$ from centimeters to
micrometers can increase the confinement frequency by orders of magnitude.
In addition, one can envision a ``lab on a chip,'' in which multiple devices are integrated on a single device,
expediting applications for complex manipulation of fermionic atoms for simulations of strongly correlated systems, quantum transport
experiments, collision-insensitive clocks, and precise interferometry \cite{qpc,anderson}.
The strong confinement provided by a microfabricated electromagnet ($\mu$EM) trap also has a practical advantage: it facilitates faster
cooling, which relaxes constraints on vacuum quality and 
leads to a tremendous simplification over traditional DFG experiments that require multiple ovens,
Zeeman slowers, or two magneto-optical traps (MOTs).

%
\begin{figure*}[!t] 
\includegraphics[width=4in]{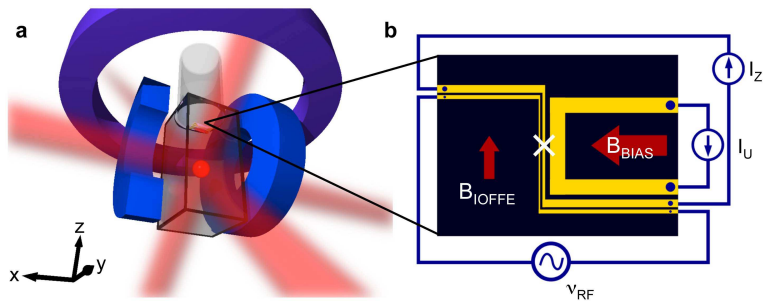}
\caption{{\bf A simple apparatus for Fermi degeneracy.} 
{\bf (a)} The dual-species MOT (red sphere) is formed at the intersection of six laser beams. 
The cloud is then magnetically trapped using external quadrupole coils (blue), transported 5\,cm vertically 
using an offset coil (purple), and compressed in the $\mu$EM trap. 
{\bf (b)} 
Schematic diagram of the central region of the $\mu$EM chip. 
A magnetic trap is formed 180\,$\mu$m above the surface at the location 
marked with a white ``X'' 
by applying $I_\mathrm{Z}=2.0$~A, $I_\mathrm{U}=30$~mA, $B_\mathrm{BIAS} = 21.4 $\,G, and 
$B_\mathrm{IOFFE} = 5.2$\,G.
Wire widths from left to right are 20\,$\mu$m, 60\,$\mu$m, and 420\,$\mu$m.}
\label{fig:apparatus} 
\end{figure*}
In our system \cite{Aubin_JLTP2005}, the entire experimental cycle takes place in a single vapour cell (Fig.~\ref{fig:apparatus}a). Counter-propagating laser beams collect, cool, and trap $2\times10^7$ $\pf$ and $10^9$ $\re$ atoms in a MOT. Atoms are transferred to a purely magnetic trap formed by external quadrupole coils 
and transported to the chip 5\,cm away. Figure~\ref{fig:apparatus}b shows several microscopic gold wires supported by the substrate. In the presence of uniform magnetic fields, current flowing through the central  `Z'-shaped wire creates a magnetic field minimum above the chip. At the centre of this trap, the $\pf$ radial (longitudinal) oscillation frequency is $\omega_\perp/2\pi = 826\pm7$\,Hz ($\omega_\ell/2\pi=46.2\pm0.7$\,Hz).
The corresponding $\re$ trap frequencies are a factor of $\sqrt{m_{Rb}/m_K} \approx 1.47$ smaller,
where $m_\mathrm{Rb}$ and $m_\mathrm{K}$ are the atomic masses of $\re$ and $\pf$, respectively.
%

%
After loading,
the 1.1-mK-deep chip trap holds approximately $2\times10^5$ $\pf$ and $2\times10^7$ $\re$ doubly spin-polarized atoms, at a temperature $\gtrsim 300$\,$\mu$K. Lower temperatures are achieved by forced evaporative cooling of $\re$. 
A transverse magnetic field oscillating at RF frequency $\nu_\mathrm{RF}$ (typically swept from $30$\,MHz to $3.61$\,MHz) selectively removes the highest energy  $\re$ atoms by driving spin-flip transitions to untrapped states.
The $\pf$ atoms, with smaller Zeeman splittings,  are not ejected but are sympathetically cooled \cite{Wieman,schreck,truscott} by thermalizing with the $\re$ reservoir via $\pf$-$\re$ collisions \cite{roati,Jin_BECDFG_PRA2004,ZurichDFG,SengstockBECDFG}. 
%
%
\setcounter{figure}{1}
\begin{figure}[!b]
\includegraphics[width=3in]{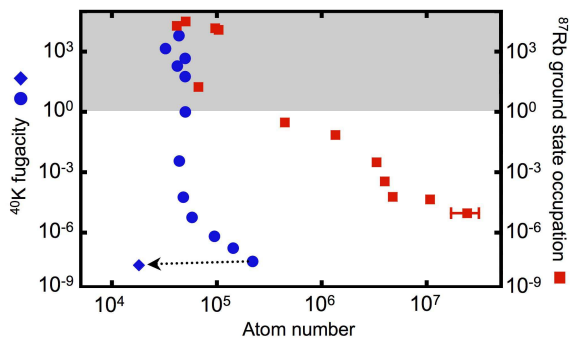}
\caption{{\bf Sympathetic cooling in a chip trap.}
Spin-polarized fermions without a bosonic bath cannot be successfully evaporated (blue diamond). However, if bosonic $\re$ (red squares) is evaporatively cooled, then fermionic $\pf$ is sympathetically cooled (blue dots) to quantum degeneracy (greyed area). For bosonic $\re$, the vertical axis is the occupation of the ground state; for fermionic $\pf$, the vertical axis is the fugacity, as discussed in the text. These two quantities are equivalent in the non-degenerate limit. A typical run-to-run spread in atom number is shown on the right-most point; all vertical error bars are smaller than the marker size.}
\label{fig:path}
\end{figure} 

The evolution of temperature $T$ and atom number $N$ during sympathetic cooling is measured by releasing atoms from the trap and observing their expansion with absorptive imaging. 
Figure~\ref{fig:path} shows the cooling of $\pf$ and $\re$  to quantum degeneracy.
In the degenerate regime, bosons accumulate in the ground state (forming a BEC), while fermions fill the lowest energy levels of the trap with near-unity occupation.
Fermi degeneracy can be quantified with the fugacity $\mathcal{Z}$: the ground state has occupation $\mathcal{Z}/(1+\mathcal{Z})$, which approaches 1 in the high $\mathcal{Z}$ degenerate limit and $\mathcal{Z}$ in the non-degenerate limit.
Due to the tight confinement of the $\mu$EM trap, cooling increases the $\pf$ fugacity by $10^{12}$ in only 6~s. The steep ascent of $\pf$ in Fig.~\ref{fig:path} also demonstrates the efficiency of sympathetic cooling. Since $\pf$ is a rare isotope and therefore more difficult to collect from vapour than $\re$, this inherent efficiency is a significant advantage over direct evaporation.
To our knowledge, this is the first observation of sympathetic cooling, of Fermi degeneracy, and of dual degeneracy in a $\mu$EM trap.

\begin{figure}[t!]
\centerline{\includegraphics[width=3.4in]{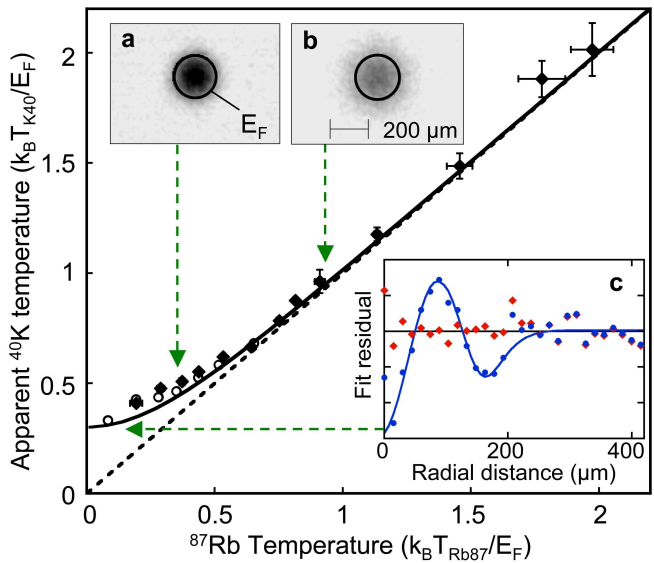}}
\caption{{\bf Observation of Fermi statistics.} Due to Pauli pressure, Fermi degenerate $\pf$ clouds appear to stop getting colder, even when the reservoir temperature approaches zero. The apparent temperature of the fermions, as measured by Gaussian fits to images of $\pf$ clouds, is plotted versus temperature of both thermal (diamonds) and Bose-condensed (circles) $\re$. Data is compared to a Gaussian fit of theoretically generated ideal Fermi distribution (solid line) and its classical expectation (dashed line). 
Both temperatures are scaled by the Fermi energy $E_\mathrm{F}$ of each $\pf$ cloud.
Error bars are statistical, with uncertainty smaller than the sizes of symbols for lower temperature data.
Absorption images are shown for $k_\mathrm{B} T/E_\mathrm{F}=0.35$ \textbf{(a)} and $0.95$ \textbf{(b)}, including a black circle indicating the Fermi energy $E_\mathrm{F}$. 
\textbf{(c)} A closer look at the fermion cloud shape reveals that it does not follow a Boltzmann distribution. The fit residuals 
of a radially averaged cloud profile show a strong systematic deviation when assuming Boltzmann (blue circles) instead of Fermi (red diamonds) statistics. A degenerate Fermi cloud is flatter at its centre than a Boltzmann distribution, and falls more sharply to zero near its edge.}
\label{fig:nonclassical} 
\end{figure}

Below $T\approx1$\,$\mu$K, we observe two independent signatures of Fermi degeneracy.
First, we compare the rms cloud size of $\pf$ and $\re$ (or its non-condensed fraction) by fitting the density profiles to a Gaussian profile. As described in Methods, this is an appropriate method for finding the temperature of a classical Boltzmann gas. Figure~\ref{fig:nonclassical} shows that the apparent (i.e., Gaussian-estimated) $\pf$ temperature approaches a finite value while the $\re$ temperature approaches zero, even though the two gases are in good thermal contact. In fact, this deviation is evidence of the  ``Pauli pressure'' expected of a gas obeying Fermi statistics \cite{truscott}: at zero temperature, fermions fill all available states up to the Fermi energy $E_\mathrm{F} = \hbar (6 N \omega_\perp^2 \omega_\ell)^{1/3}$, where $N$ is the number of fermions.
For our typical parameters,  $E_\mathrm{F} \approx k_\mathrm{B} \times $ 1.1\,$\mu$K.
We plot data with thermal (diamonds) and Bose-condensed (circles) $\re$ separately,
to show that the density-dependent attractive interaction between $\pf$ and $\re$ does not affect the release energy significantly.
A second signature of Fermi statistics is evident in the shape of the cloud. Figure~\ref{fig:nonclassical}c compares the residuals of a Gaussian fit (which assumes
Boltzmann statistics) to the residuals of a fit which assumes Fermi-Dirac statistics. The Fermi distribution describes the data well, with a $\chi^2$ three times lower than the Gaussian fit. After all $\re$ atoms have been evaporated, we use Fermi-Dirac fits to 
measure temperature, and find $k_\mathrm{B} T/E_\mathrm{F}$  as low as $0.09\pm 0.05$ with as many as $4\times 10^4$ $\pf$ atoms.

\begin{figure}[t!]
\includegraphics[width=3.25in]{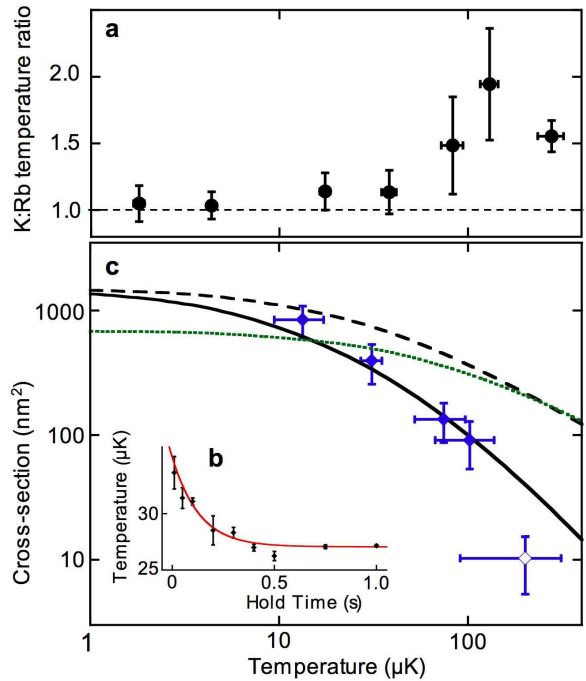}
\caption{{\bf Cross-species thermalisation.} {\bf (a)} The ratio of the temperature of $\pf$ to the temperature of $\re$ approaches unity as $\pf$ temperature is lowered during sympathetic cooling. {\bf (b)} We measure cross-thermalisation by abruptly reducing the temperature of $\re$ and watching the temperature of $\pf$ relax versus time. The data shown has an asymptotic $\pf$ temperature of 27\,$\mu$K.  {\bf (c)} Measurements of $\sigma_\mathrm{KRb}$ (diamonds) are compared to the ``naive'' model (dashed) and an effective range model (solid), both described in the text. The error bars shown in all parts of the figure are statistical, with the exception of temperature error bars in (c), which show the spread between initial and final $\pf$ temperature. The highest temperature point (open diamond) did not completely thermalise, and is discussed further in the text.
For reference, the s-wave $\sigma_\mathrm{RbRb}$ is also shown (dotted).
}
\label{fig:thermalisation} 
\end{figure}

We empirically optimize the sympathetic cooling trajectory, 
and find that RF sweep times faster than 6~s are not successful, whereas $\re$ alone can be cooled to degeneracy in 2~s. This 
indicates that $\pf$ and $\re$ rethermalise more slowly than $\re$ with itself. 
Measuring the temperature 
ratio during sympathetic cooling (Fig.~\ref{fig:thermalisation}a) reveals that $\pf$ lags behind $\re$ at high temperatures, despite the fact that our optimal frequency ramp 
starts slowly (when the atoms are hottest), and accelerates at lower temperatures.

In the low-temperature limit, we do not expect the cross-species thermalisation to lag the $\re$-$\re$ thermalisation, since the $\pf$-$\re$ cross-section $\sigma_\mathrm{KRb}=1480\pm70$\,nm$^2$ \cite{ferlaino} exceeds the $\re$-$\re$ cross-section, $\sigma_\mathrm{RbRb}=689.6\pm0.3$\,nm$^2$ \cite{verhaar}. However, several conflicting values for $\sigma_\mathrm{KRb}$ have been recently presented
\cite{Inguscio_Science2002,Jin_BECDFG_PRA2004,inouye,%
SengstockBECDFG,ferlaino}.

To investigate $\sigma_\mathrm{KRb}$ further, we 
measure the cross-species thermalisation rate \cite{ArimondoPRA2005} at several temperatures. Starting from equilibrium, 
we abruptly cool $\re$ by reducing $\nu_\mathrm{RF}$, wait for a variable hold time to allow cross-thermalisation, and then measure the $\pf$ temperature, as shown
in Fig.~\ref{fig:thermalisation}b.  We repeat 
this measurement at several temperatures, and fit each to the model of 
Ref.~\cite{Grimm_Thermalization_ApplPhysB2001}. 
We find that the cross-section has a dramatic dependence on temperature (see Fig.~\ref{fig:thermalisation}c), decreasing over an order of magnitude between 10 and 200\,$\mu$K. 
%
%
%
%

The simplest model for atom-atom scattering uses a delta-function contact potential. 
Fig.~\ref{fig:thermalisation}c shows that the s-wave scattering cross-section of this ``naive'' model (further described in Methods) would predict a higher $\sigma_\mathrm{KRb}$ than $\sigma_\mathrm{RbRb}$ throughout our cooling cycle, in stark contrast to our data. Better agreement is given by an effective range model \cite{ScattTheory}, which includes a reduction in scattering phase (and thus cross-section) below the naive expectation. Our highest temperature data point lies below the effective range prediction, 
however a more sophisticated analysis may be required to extract a quantitative measurement for this point, due to severe trap anharmonicity at high temperature. Overall, both data and theory show that the $\pf$-$\re$ cross-section is reduced well below the $\re$-$\re$ cross-section for a large range of temperatures, explaining the requirement for a slow initial RF frequency sweep for sympathetic cooling.  Below $20$~$\mu$K, where no temperature lag is observed, $\sigma_\mathrm{KRb}$ exceeds $\sigma_\mathrm{RbRb}$.

We attribute the observed reduction in scattering cross-section to the onset of the Ramsauer-Townsend effect, in which the s-wave scattering phase and cross-section approach zero for a particular value of relative energies between particles \cite{Townsend}. At higher temperatures, the scattering cross-section should increase again, however free evaporation from our trap limits our measurements to below 300\,$\mu$K. Additional partial waves may also affect scattering above the p-wave threshold of 110\,$\mu$K. Despite the high-temperature reduction in cross-section, 
$\pf$ and $\re$ remain relatively good sympathetic cooling partners.
Recent measurements  of $\re$-$^{6}$Li sympathetic cooling \cite{ZimmermanRbLi}, for instance, suggest a zero-temperature cross-section approximately 100 times smaller than $\sigma_\mathrm{KRb}$, i.e., a maximum cross-section roughly equal to the lowest value we measure here.

The high collision rates in mixtures trapped with a $\mu$EM enable us to
cool fermions sympathetically to quantum degeneracy in 6\,s, faster than previously possible.
For comparison, {\em direct} evaporation of optically trapped $^6$Li to Fermi degeneracy has been achieved in 3.5\,s \cite{ohara}, close to a Feshbach resonance.
However, the all-optical approach does not lend itself easily to sympathetic cooling.
The lower efficiency of direct evaporation requires a stronger source of $\pf$, previously only achieved with multiple vacuum chambers \cite{Jin_Science1999}, 
due to low available isotope abundance.

In conclusion, we have achieved simultaneous quantum degeneracy of bosonic and fermionic atoms in a $\mu$EM trap and demonstrated an approach that can simplify future research with cold fermions.
One prospect is the observation of Pauli blocking in light scattering off degenerate fermions \cite{PauliBlockLight, ExploringDFG}.
The high $\mu$EM trap frequencies boost the ratio of Fermi energy $E_\mathrm{F}$ to the recoil energy $\hbar^2 k^2 /2m_\mathrm{K}$ to $\sim$2.5, within the range necessary to explore such quantum optical effects.

\section*{Methods}
{\bf Loading.} Our experimental cycle is similar to that described in Ref.~\cite{Aubin_JLTP2005}, with several key modifications emphasized here and in the main text.
Approximately 600\,mW of incoherent 405\,nm light  
desorbs $\re$ and $\pf$ atoms from the pyrex vacuum cell walls, boosting the MOT
atom number one hundred-fold compared to loading from the background vapour. Potassium alone is first loaded into the MOT for 25 s, after which $\re$ is loaded for an additional 3-5 s, while maintaining the $\pf$ population.
Both MOTs operate with a detuning of $-26$\,MHz, until the last 10\,ms, when $\pf$ is compressed with a $-5$\,MHz detuning.
After MOT loading, 3\,ms of optical molasses cooling is applied to
the $\re$ atoms, while the $\pf$ atoms are optically pumped into the $\ket{F=9/2,m_F = 9/2}$ hyperfine ground state.

{\bf Micro-electromagnet trap.} 7-$\mu$m-thick gold wires are patterned lithographically and electroplated on a silicon substrate. Two defects are present near the centre of the principal Z-wire,
which result in the formation of three ``dimples'' in the trapping potential. We use the magnetic gradient generated by 30~mA of current through the U-wire to centre the magnetic trap on one of these dimples.

{\bf Fitting absorption image data.} Degenerate Fermi clouds are fit using a semi-classical expression for the optical density: 
$A f_2(\mathcal{Z} \exp{[-\varrho^2 / 2 r^2]}),$ where $\varrho$ is the radial coordinate, $A f_2(\mathcal{Z})$ 
is the peak optical
density, $\mathcal{Z}$ the fugacity, and $f_{\nu}(q) = -\sum_{\ell=1} ^{\infty} (-q)^\ell/\ell^{\nu}$ is the
Fermi-Dirac function.  The temperature is given by $k_\mathrm{B} T = r^2 m_\mathrm{K} / (\omega_\perp^{-2} + t^2),$ where $r$ is the fit width and $t$ the time of flight.  Atom
number is extracted using $N = 2\pi r^2 f_3 (\mathcal{Z}) A/\sigma_{\lambda}$, where
$\sigma_{\lambda}$ is the resonant absorption cross-section.  $T/T_\mathrm{F}$ can be extracted directly from the fugacity using
$(T/T_\mathrm{F})^{-3} = 6 f_3(\mathcal{Z})$. Non-degenerate clouds are fit to a Gaussian distribution $A \exp{[-\varrho^2 / 2 r^2]}$, with the same interpretation of $r$.
Probes along both $\hat{x}$ and $\hat{y}$  (see Fig.~\ref{fig:apparatus}) were used for imaging. Comparison of temperature measurements along axes of expansion suggest a 20\,nK  kick (possibly magnetic) is given to clouds along $\hat{z}$, and that other temperatures agree systematically at the 5\% level.
Data for residuals shown in Fig.~\ref{fig:nonclassical}c 
is radially averaged about an ellipse defined by the two trap frequencies of the image plane.  This 1D radial data set is binned into 2-pixel bins, and fit as described.

{\bf Scattering theories.} The ``naive'' interaction model discussed in the text gives $\sigma_\mathrm{KRb} = 4 \pi a^2 / (1+a^2 k^2)$, where $a$ is the s-wave scattering length and $k$ is the relative wave vector in the centre of mass frame. Fig.~\ref{fig:thermalisation}c shows the thermally averaged theory curves. Including the next order correction in the s-wave scattering amplitude 
$f(k)=-[1/a + i k + k^2 r_\mathrm{e}/2 + \ldots]^{-1}$ 
requires an effective range, which we calculate using 
\cite{ScattTheory} to be $r_\mathrm{e}=20.2\pm0.3$\,nm, for $a_\mathrm{KRb}=-10.8 \pm 0.3$\,nm \cite{ferlaino}. 

{\bf Analysis of thermalisation data.} When the $\re$ atom number $N_\mathrm{Rb}$ is much larger than the $\pf$ atom number, the relaxation of the $\pf$ temperature $T$ to $T_\mathrm{Rb}$ is 
described by 
$\dot{u} = -u  \tau^{-1} (1 + m_\mathrm{Rb} u/(m_\mathrm{Rb}+m_\mathrm{K}) )^{\frac{1}{2}} ( 1 + u/2)^{-\frac{3}{2}},$
where $u \equiv (T/T_\mathrm{Rb})-1$, and thermalisation time $\tau$ given by 
\begin{equation}
\frac{1}{\tau} = \frac{\sqrt{2}}{3 \pi^2} \frac{\sigma_\mathrm{KRb}}{k_\mathrm{B} T_\mathrm{Rb}} \frac{\sqrt{m_\mathrm{K}} m_\mathrm{Rb}^2 \omega_\perp^2 \omega_\ell}{\left(m_\mathrm{K} + m_\mathrm{Rb}\right)^{3/2}} N_\mathrm{Rb},
\end{equation}
in which trap frequencies are for $\re$ \cite{Grimm_Thermalization_ApplPhysB2001}. Fitting for $\tau$ allows us to extract $\sigma_\mathrm{KRb}$. Note that all thermalisation data is taken with $N_\mathrm{K}$ below 4\% of $N_\mathrm{Rb}$. 

Data in Fig.~\ref{fig:thermalisation}c is analysed assuming a temperature-independent cross-section within the range of initial to final temperature. To check this assumption, we re-analyse the data using a self-consistent method that assumes an effective range temperature-dependence, and find a small upward shift of the best fit cross-section values. Using this shift as an estimate of the methodology-dependent systematic error,  
we fit our four lowest temperature measurements with the 
effective range model, and find $a_\mathrm{KRb}= -9.9 \pm 1.4 \pm 2.2 $\,nm, in agreement with Ref.~\cite{ferlaino}. The second uncertainty reported is systematic, and also includes uncertainty in the $\re$ number calibration.

{\bf Acknowledgments:} We would like to thank D.\ Jin, J.\ Dalibard, J.\ Bohm, and 
D.\ Guery-Odelin for helpful conversations about scattering theory, and A. Simoni for sending us unpublished $\pf$-$\re$ cross-section calculations. We also thank
N.\ Bigelow, A.\ Aspect, T.\ Schumm, and H.\ Moritz for stimulating conversations,
P.\ Bouyer and R.\ Nyman for providing a tapered amplifier used in this work,
and J.\ Est\`eve for fabricating the chip used in this work. 
This work is supported by the NSERC, CFI, OIT, PRO, CRC, and 
Research Corporation. S.A., L.J.L. and D.M. acknowledge support from NSERC. 
M.H.T.E. acknowledges support from OGS.
{\bf Competing financial interests:}
The authors declare that they have no competing financial interests.

\end{document}